# THE CASE FOR DYNAMIC KEY DISTRIBUTION FOR PKI-BASED VANETS

Ahmed H. Salem[1], Ayman Abdel-Hamid[2], and Mohamad Abou El-Nasr[3]

[1]Computer Science Department, Old Dominion University, Norfolk, VA, USA
[2]College of Computing and Information Technology, Arab Academy for Science, Technology and Maritime Transport, Alexandria, Egypt
[3]College of Engineering and Technology, Arab Academy for Science, Technology and Maritime Transport, Alexandria, Egypt

## ABSTRACT

*Vehicular Ad hoc Networks (VANETs) are becoming a reality where secure communication is a prerequisite. Public key infrastructure (PKI) can be used to secure VANETs where an onboard tamper proof device (TPD) stores a number of encryption keys which are renewed upon visiting a certificate authority (CA). We previously proposed a dynamic key distribution protocol for PKI-based VANETs [1] to reduce the role of the TPD. A vehicle dynamically requests a key from its nearest road side unit. This request is propagated through network infrastructure to reach a CA cloud and a key is securely returned. A proposed key revocation mechanism reduced the number of messages needed for revocation through Certificate Revocation List (CRL) distribution. In this paper, performance evaluation and security of the proposed dynamic key distribution is investigated analytically and through network simulation. Furthermore, extensive analysis is performed to demonstrate how the proposed protocol can dynamically support efficient and cost-reduced key distribution. Analysis and performance evaluation results clearly make the case for dynamic key distribution for PKI-based VANETS.*

## 1. INTRODUCTION

Vehicular Ad-Hoc Networks (VANETs) are a very promising evolution of Mobile Ad-Hoc Networks (MANETs) because of its capability in solving road problems such as traffic congestion and vehicular safety. A VANET consists of network entities, including On board units (OBU), Road side units (RSU), and a Certificate Authority (CA). These entities form a network hierarchy as illustrated in Fig. 1. Either a car manufacturer or government authority plays the role of the CA responsible for key generation and distribution. The RSUs are responsible for moderating communication between vehicles, and delivering messages from the CA. The RSUs are installed in standalone towers and organized according to the network topology. The OBUs are loaded in vehicles and are responsible for receiving and validating messages [2].

Vehicle-to-Infrastructure (V2I), Vehicle-to-Vehicle (V2V), and Infrastructure-to-Infrastructure (I2I) are vehicular communication modes as shown in Fig. 2. V2I allows vehicles to communicate with roadside units, while V2V allows vehicles to communicate with each other. I2I allows roadside units to communicate with each other [2].

For security key storage, a VANET depends on a Tamper Proof Device (TPD) as the main storage for security communication elements [2]. Vehicles, while making their periodical visits to the CA, acquire some security elements and are stored in both the vehicle and the CA [3]. Public Key Infrastructure (PKI) was proven to be secure and efficient for the use in VANETs [4]. However, the classical use of PKI involved preloading the vehicle with a large number of keys that are vulnerable to attacks.

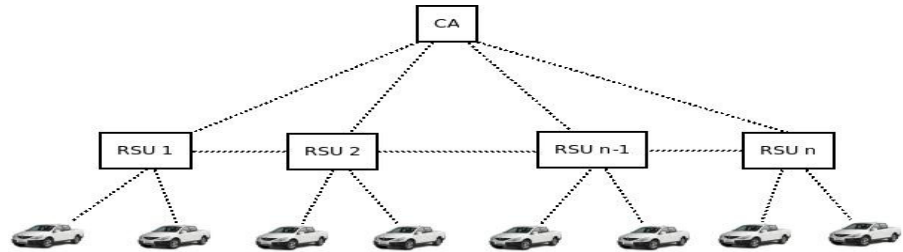

Figure 1. Standard VANET Hierarchy

In an earlier work [1], we proposed a dynamic key distribution protocol eliminating the need to store large amount of keys for PKI support. The role of the TPD is reduced to carrying the vehicle's unique information used as a secret key between the CA and the vehicle. This unique information is the Electronic license Plate (ELP), Electronic chassis number (ECN), and both will form the Vehicle Authentication Code (VAC). The VAC is used to dynamically authenticate the vehicles supporting PKI-based VANETs. An RSU manager is introduced to the standard VANET hierarchy (Fig. 1) to improve message passing as will be later clarified in section 3. In this paper, further details of the proposed key distribution protocol [1] are presented with in-depth security analysis while focusing on security attacks' resistance. Performance evaluation experiments clearly highlight the feasibility and scalability of the proposed protocol and strongly make the case for dynamic key distribution for PKI-based VANETs.

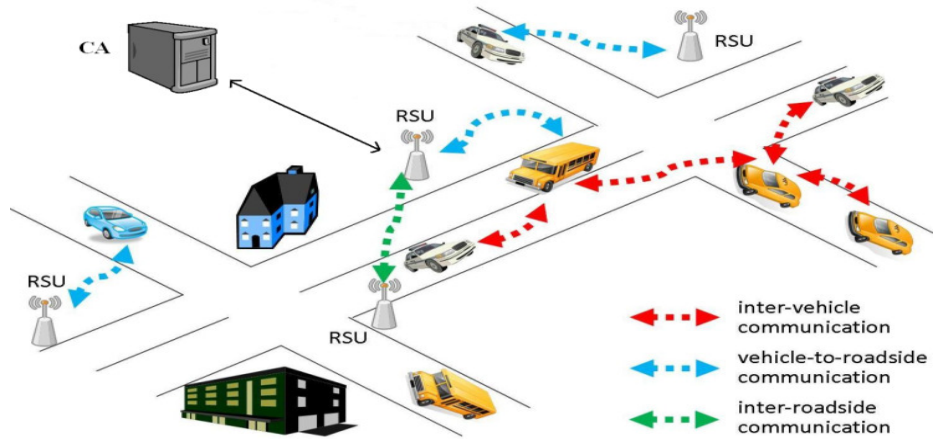

Figure 2. VANET nodes and communication Techniques

The rest of this paper is organized as follows. Section 2 surveys related work. In Section 3, the dynamic key distribution protocol, assumptions, workflow, and countermeasures against possible security attacks are presented. Performance evaluation through analytical analysis and network simulation experiments are presented in Section 4. Section 5 concludes the paper and discusses future work.

## 2. RELATED WORK

VANETs can be used as an infrastructure for many applications. These applications can be classified into four classes as follows [2]. *General Information Services*: includes mobile Internet and advertising; *Information Services*: for vehicle safety which includes warning messages, road awareness, and condition alerts; *Individual motion control*: using inter-vehicle communication, which includes collision avoidance, and cruise control; and *Group motion control*: using inter-vehicle communication that includes optimized path planning, and rights based traffic control.

The attacks against VANETs can be classified as follows [3]. *Bogus information*: attackers flood the network with incorrect information to affect other driver's behavior; *Cheating with positioning information*: where attackers change their position, speed, and direction to escape from liability; *ID disclosure*: to track the location of vehicles; and *Denial of Service*: to bring down the network or cause an accident; and *Masquerade*: where an attacker pretends to be another vehicle using fake identities.

Efficiently dealing with the key distribution problem is a cornerstone in the design of security algorithms. In previous research efforts, an authority is responsible for key issuance whether a vehicle maintenance center or a license renewal center. Any of these authorities can generate keys and save them in a Hardware Security Module (HSM) [6] or TPD [11]. The information stored on the TPD should not be accessible to anyone but trusted authorities. This device is considered the main component of the key management solution.

Hubaux et al. [5,6] used the TPD as a storage medium for the public key of the certificate authority. In this approach, a node sends to another the hash of the global key. If it is the same as the hash generated from the key stored in the TPD, the communication is established between these two nodes. This approach relies on the TPD, which is expensive, and non-reliable device.

Aslam and Zou [13] tried to ignore the TPD as a non-realistic device and used an on-board device loaded with prepaid cards containing the keys. The cards contain an identification and certificate. During initialization, the user information will be maintained with the provider and not stored in the device. When a user enters a service area, he makes the service payment using the on-board payment device. The message is encrypted by the provider's public key, thus hiding the device certificate and services requested from eavesdroppers. The user is issued a pseudonym that will be valid for a given period/area. This is nearly the behavior using TPD.

Zhang [14] assumed that a vehicle registers itself with a public/private key generator. When it enters the communication area of an RSU, it initiates a mutual authentication process with the RSU. Diffie-Hellman algorithm is used to exchange a symmetric key. RSU and other vehicles in range receive an encrypted message and a message authentication code (MAC). The MAC is generated based on the message and symmetric key shared with the RSU. Only the RSU can validate that the MAC as it is the only other owner of the symmetric key. If the RSU validates the MAC, it sends a valid message to the vehicles. To avoid the case of RSU failure, V2V communication will be performed instead ignoring V2I. PKI is used and a TPD is used for key storage. This approach ensures privacy but with no communication to the certificate authority to ensure the vehicle is legitimate. In addition, it treats each RSU as a separate network with no supposed mean of communication and cooperation. This will make the vehicle repeat this authentication process every time it enters the communication range of an RSU.

Wasef and Shen [15] introduced a certificated distribution protocol for VANETs. They used a four level architecture consisting of Master CA (MA), CA, RSU, and OBU. The difference from the classical VANET in Fig. 1 is adding the MA that is the highest level of system security and trust. The authors use standard PKI and certificate revocation list (CRL) broadcasting between all levels [16]. They use two CRLs for RSUs and OBUs respectively and GPS to locate vehicles. This will flood the network with useless information.

Hao et al. [17] introduced a distributed key management reducing the computation overhead. The authors used group communication between vehicles. The vehicle is supposed to periodically resend its location to the RSU. This will lead to problems concerning group leader and vehicles joining and leaving the group.

Laberteaux et al. [18] introduced a secure CRL distribution protocol depending on V2V communication which results in 99% network coverage when compared to 91% achieved by V2I CRL distribution. Although this work achieves a very high percentage in CRL distribution it affects network utilization by using such network flooding protocol.

One of the principles of the public key infrastructure (PKI) that should be handled effectively is key revocation, and how to distribute a certificate revocation list (CRL) among the VANETs efficiently without loading the network. Use of a CRL has many problems concerning VANETs. The CRL will be very long due to large number of vehicles. Short life time certificates will also create a vulnerability window. The infrastructure might not support a CRL especially at the start of deployment.

Hubaux et al. [5, 6] introduced three solutions for CRL distribution problems. The Revocation Protocol of the Tamper Proof Device (RTPD), Revocation Protocol using Compressed CRL (RCCRL), and Distributed Revocation Protocol (DRP). In RTDP, the CA sends an encrypted message to a vehicle to erase all keys in the TPD. The CA should know the vehicle location or the most recent one from a location database. The CA should receive an ACK from the vehicle. If no ACK is received, the CA sends the message via FM Radio or satellite. In RCCRL, Bloom filters are used to reduce the CRL size to be few kilobytes then broadcast them. In DRP, the vehicles accumulate accusations against misbehaving vehicles and then are reported to the CA.

Aslam and Zou [13] introduced some precautions regarding revocation. First, the CA checks for revoked vehicles keys before key issuance to check that the user didn't revoke the vehicle's key. Second, CRL distributed within the RSU range. Third, Key provider companies can limit their work in specific geographical areas.

Papadimitratos et al. [6] proposed a revocation protocol that divides the CRL into M equal pieces. The M pieces are encoded using erasure encoding into N redundant pieces. Each piece has a header then signed by the CA's private key. The header contains CRL version, time stamp, sequence number, and CA's Id. The pieces are sent to the RSU for broadcasting. When a vehicle receives a piece, it verifies the time stamp then the signature. If valid, the vehicle verifies if it has the piece already stored or not. After having enough pieces, it decodes the CRL. In this algorithm, there are multiple drawbacks. First, the encoding used avoids missing some pieces during sending but will result in flooding the network with unneeded packets. In addition, broadcasting to the whole network will result in decreasing the network performance.

Nowatkowski [20] proposed two protocols to optimize the CRL distribution overhead. The first is Most Pieces Broadcast (MPB). MPB selects the node with maximum CRL pieces to broadcast. The second protocol is Generation Per Channel (GPC). GPC splits the CRL into multiple parts. Each one is sent on a different channel. The OBU downloads CRL pieces from every channel with equal probability. This results in the download completion for all pieces approximately at the same time. These protocols overload the network in addition to the difficulty of allocating multiple channels required by GPC.

# 3. DYNAMIC KEY DISTRIBUTION

This section explains the proposed network model, the assumptions on which the protocol is built, how the key distribution works, how to deal with vehicle's movement, how revocation is handled, resistance to security attacks, and how the proposed protocol compares to related work.

## 3.1 Network Model

The proposed solution introduces an RSU manager to the standard VANET hierarchy as illustrated in Fig. 3. The RSU manager stores the RSU locations and forwards CA messages to a specific RSU. The RSU manager is assumed to have all information regarding the RSUs under its authority. The RSU manager is responsible for keeping track of the current RSUs that requested keys for each vehicle until that key expires. A group of RSUs and their areas of responsibility form an administrative domain. This domain represents a suitable geographical division as dictated by civic authorities. To generalize, a CA cloud is assumed to be deployed. Augmenting the CA to become a CA cloud will make the RSU manager free to communicate with such cloud without having any limitation of a down server or unauthorized region. This could be achieved for example by assigning an anycast address to the CA cloud.

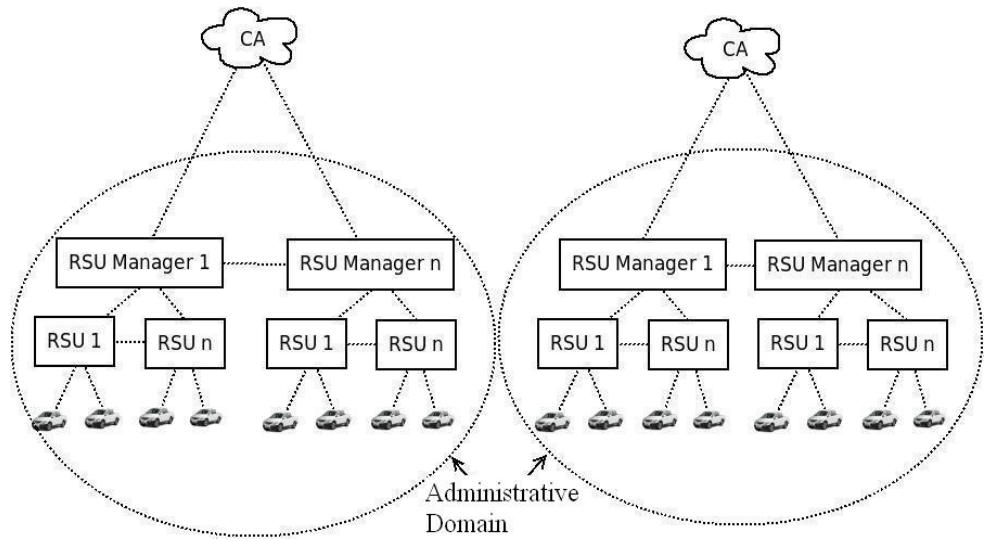

Figure 3. Proposed VANET Hierarchy

## 3.2 Assumptions

The following assumptions are in effect for the proposed protocol:

1- Vehicles are free to communicate with each other after authentication.
2- RSUs have a storage medium to store state about the vehicles passing by them as long as their certificate lifetime did not expire.
3- RSU manager is responsible for a group of RSUs to cover the whole network as shown in Fig. 3.
4- RSU manager is responsible for locating RSUs and secure messaging between vehicles in their area of responsibility.
5- Soft handover is supported between adjacent RSUs to ensure message passing while the vehicle moves on.
6- Vehicles have a unique authentication key called vehicle authentication code (VAC) which is composed of the chassis number and electronic license plate (ELP) and is known only by the vehicle and the CA.

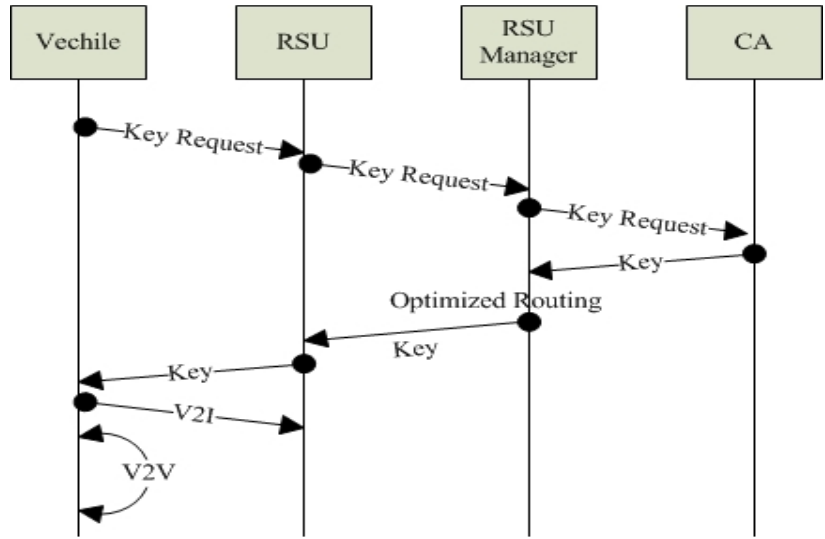

(1) Vehicle → RSU:      ELP ‖ N1
(2) RSU → RSUM:      $E_{RSUM-KU}( E_{RSU-KR}(ELP ‖ N1 ‖ N2 ) )$
(3) RSUM → CA:      $E_{CA-KU}( E_{RSUM-KR}(ELP ‖ N1 ‖ N2 ‖ N3) )$
(4) CA → RSUM:      $E_{RSUM-KU} (E_{CA-KR}(E_{VAC}( key ‖ f(N1)) ‖ N2 ‖ N3))$
(5) RSUM → RSU:      $E_{RSU-KU} (E_{RSUM-KR} (E_{VAC}( key ‖ f(N1)) ‖ N2 ))$
(6) RSU → Vehicle:      $E_{VAC}( key ‖ f(N1))$

**RSUM**: RSU manager        **KR**: Private Key        **f**: Agreed upon function
**KU**: Public Key        **N1, N2, N3**: nonces

Figure 4. Messaging sequence to acquire a key

## 3.3 Key Distribution Workflow

Fig. 4 depicts the proposed protocol workflow. The OBU sends a request to the RSU to acquire network authentication. The authentication scenario will be as follows. A vehicle requesting authentication sends its ELP with a nonce N1 to the nearest RSU. The RSU will append its nonce and double encrypts the ELP and nonces N1 and N2 with the RSU's private key and RSU manager's public key, respectively then sends to the RSU manger. The RSU manager checks the authenticity and integrity of the received message. If the message checks out, the RSU manger appends its nonce N3 then double encrypts the ELP and nonces N1, N2, and N3 with the manager's private key and the CA's public key, respectively then passes it to the CA. The CA checks the authenticity and integrity of the received message. If the message checks out, the CA checks if the vehicle is black-listed or has a previous revocation request. The CA generates the key and encrypts the (key and N1) with the VAC and appends to it nonces N2 and N3 and sends it to the RSU manager which in turns checks the message. If the message checks out, it passes it to the RSU after removing its nonce. The RSU unicasts the encrypted (key and N1) after removing its nonce N2. The only vehicle that can decrypt the message is the one requesting authentication. Messages [2-5] offer authentication and confidentiality services to recipients. In future work, techniques to make the VAC more challenging to attackers will be investigated. For simplicity, we refer to a distributed *key* in Fig. 4, where in practice the distributed key material would be a PKI certificate [9].

TABLE 1
Notations and Description

| | Description |
|---|---|
| $l$ | certificate lifetime (sec) |
| $v$ | vehicle's speed (km/h) |
| $d$ | distance between two adjacent RSUs (m) |
| $r$ | radius of the area where the vehicle is most probably in (m) |
| $N$ | The number of RSUs in the whole network |
| $n$ | The number of RSUs in the circular area with radius $r$ |
| $m$ | Max. no. of messages to revoke vehicle's key |
| $t_{E2E}$ | time between revocation initiation and the actual revocation (sec.) |
| $t_{p.CA}$ | Revocation message processing time at CA (sec.) |
| $t_{CA}$ | Revocation message travel time from CA to RSU Manger |
| $t_{p.man}$ | Revocation message processing time at RSU Manager (sec.) |
| $t_{man}$ | Revocation message travel time from RSU Mgr to stored RSU (sec.) |
| $t_{p.RSU}$ | Revocation message processing time at RSU (sec.) |
| $t_{RSU}$ | Revocation message travel time between two adjacent RSUs (sec.) |
| $p$ | Percentage of RSUs used to send a message |

This can be used in applications such as advertising where the RSU manager can send the RSU suitable advertisements for that region to publish to passing vehicles, e.g., restaurants in that region. In addition, the authentication center can send warning messages to the RSU to publish to the whole VANET.

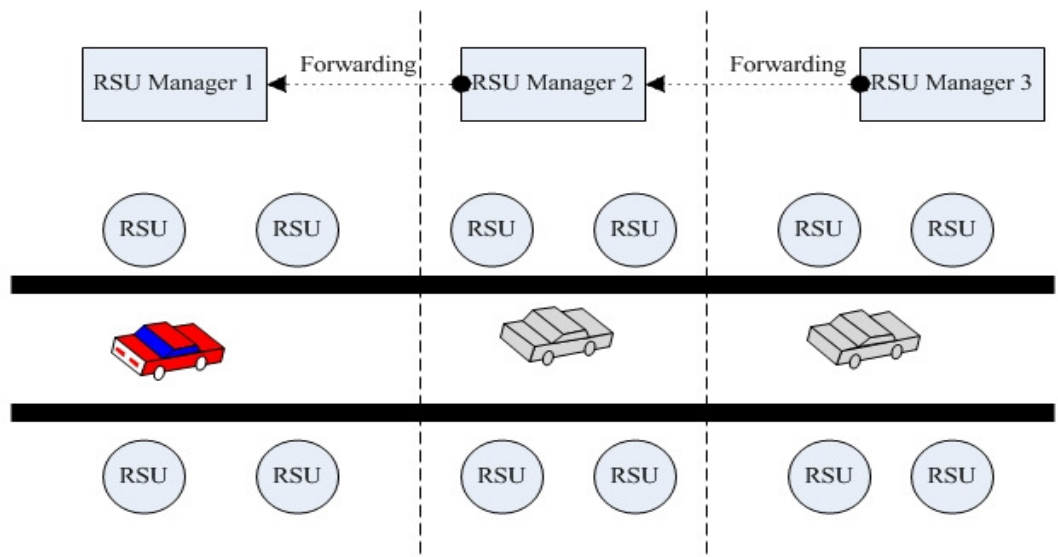

Figure 5. Sample Mobility Management Scenario

### 3.4 Mobility Management

The movement of vehicles between RSUs will trigger sending messages to the RSU manager to track the RSUs that each vehicle encountered during the certificate validity time as shown in Fig. 5. If a vehicle moved between two RSUs under different RSU managers, the target RSU is responsible for registering the vehicle with its manager. Given that the certificate did not expire, the new manager informs the previous manager that the vehicle is now under its authority. Hence, any revocation message for the vehicle should be passed to the new RSU manager. If the vehicle moved further within the same administrative domain, a chain of RSU Managers will be formed. However, as the number of the RSU managers within an administrative domain is probably low, the probability of having a long forwarding chain is low. However, this issue will be considered in future work by introducing a root manager as a layer above RSU managers to handle switching between RSU managers. The CA will keep track of RSU managers that requested keys as long as the keys are valid. The RSU manager will keep track of the RSUs and RSU managers that each vehicle visited to be used in revocation.

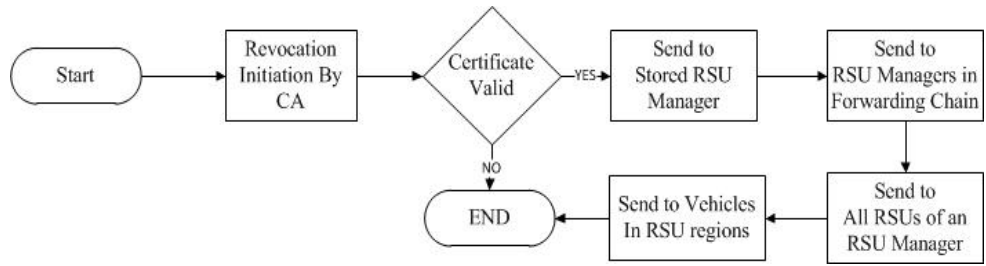

Figure 6. Certificate revocation workflow

### 3.5 Certificate Revocation

The role of the revocation protocol is to send a message to erase the certificate from the vehicle, and to warn other vehicles from dealing with the revoked-key vehicle. In previous work, CRL distribution depended on broadcasting either for the whole CRL at once or dividing it into compressed parts then broadcasting them [4].

The proposed solution depends on benefiting of the knowledge from the key distribution phase. For example, if the CA receives a request to revoke a specific vehicle's key (one possible revocation scenario). It checks if its certificate lifetime has expired or not. If it is still valid, the CA sends a revocation message to the stored RSU Manager. The RSU Manager checks if the vehicle is still registered and sends to the corresponding RSU that requested a key for the vehicle. The RSU will send the revocation message to all the vehicles in its region to warn them from dealing with this vehicle as depicted in Fig. 6.

Fig. 7 depicts the analytical parameters for revocation. The notations used throughout the paper are shown in Table 1. Depending on the vehicle's mobility scenario; its maximum speed can be deduced. For example, on a highway, a vehicle might have a maximum speed of 120 km/h. Within a city, the vehicle cannot exceed a speed of 60 km/h. Aided with the lifetime of the certificate and the original location; the maximum horizontal distance the vehicle could have crossed can be calculated. Unknowing the direction where the vehicle moved in, a circular region with radius $r$ of the horizontal distance is assumed as given by equation (1).

$$\text{(1)}$$

A vehicle moves in one direction at a time. Hence, from the circular area with radius $r$, the vehicle can only move a distance of $r$ which limits the number of RSUs that the vehicle passes by to $m$ as given by equation (2) which is equivalent to the number of required messages to revoke the key of a specific vehicle.

$$m = (r/d) + 1 \qquad\qquad (2)$$

The percentage of nodes $p$ used can be calculated using the number of RSUs used $m$ and the total number of RSUs $N$ as given by equation (3).

$$p = (m/N) * 100 \qquad\qquad (3)$$

The time interval between the revocation initiation and the actual revocation will be the sum of the CA processing time, the time the message takes to travel between the CA and the RSU manager, the processing time at the RSU manager, the message travel time between the RSU manager and the RSU, the time taken for processing at the RSUs, and the travel time the message takes between the RSUs (Equation 4).

$$t_{E2E} = t_{p.CA} \; + \; t_{CA} + \; t_{p.man} + \; t_{man} + n * (t_{p.RSU} + t_{RSU}) \qquad\qquad (4)$$

Equation 4 has some significant components that influence the result of the equation and others that are negligible. Processing times are mostly of negligible values that can even be ignored as machines' processing capabilities are continuously on the rise.

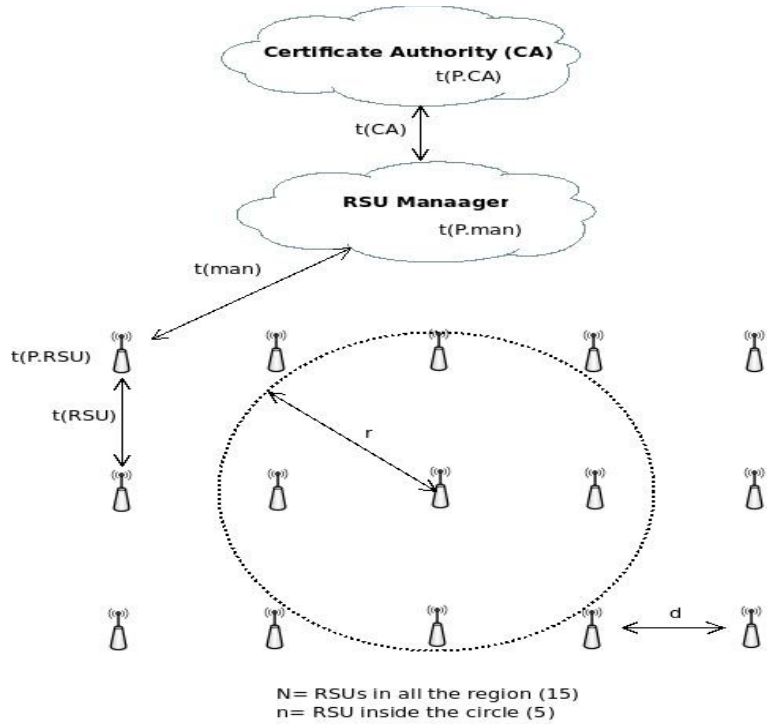

Figure 7. Revocation analytical parameters

### 3.6 Security Characteristics and Resistance to Attacks

The proposed protocol's design provides built-in countermeasures for potential security and data integrity threats. In this section, the proposed protocol's security characteristics and resistance to attacks is qualitatively outlined. Potential attacks include man-in the- middle, Sybil, and replay attacks briefly summarized as follows. *Man-in-the-middle attack*: a security attack where an intruder captures messages exchanged between two parties. Then, he generates messages based on them pretending to be an authorized member of the communication [10]. *Sybil attack*: users pretend to be others by faking their IDs [10]. *Replay attack*: is a network attack where an entity sends a prerecorded message (previously captured) to receive a data or connection that he is not authorized to access. Furthermore, a secure system must provide some services such as non-repudiation where a sender/receiver cannot possibly deny sending/receiving a specific message [10]. VANETs as most mobile systems are prone to a plenty of security threats. Raya and Hubaux [3], Wasef and Shen [4], and Du and Zhu [8] mentioned some requirements for a VANET to be secure. It should provide authentication, availability, non-repudiation, and being effective in real-time systems. In the following subsections, resistance to security attacks is qualitatively highlighted using a scenario-based approach. As terminology, (I) will denote the intruder and (V) will denote the vehicle requesting a key.

### 3.6.1 Non-repudiation and Masquerade

The proposed protocol requires the vehicle to send its ELP and then receives the key encrypted using the VAC. Since the VAC is not known except for the vehicle and CA, this will eliminate the vulnerability against non-repudiation as per the following scenario:

- Vehicle (V) is requesting a key.
- No one can send the request on behalf of (V) as no entity other than (V) and the CA know the ELP (non-repudiation).
- On the other hand, if (I) pretends to be the CA and sends a key to (V). (I) will not be able to use the key, as the VAC is required for message decryption (masquerade).

### 3.6.2 Man-in the-Middle Attack

The proposed protocol ensures that messaging between the networks nodes is confidential to the message intended receiver. This will eliminate the vulnerability to the Man-in-The-Middle attack as per the following scenario:

- (V) will send a key request to RSU with the following message (ELP‖N1).
- RSU will forward the key request till it receives a key back as previously explained in Fig. 4.
- (I) will capture the message $E_{VAC}$(key ‖ f(N1))
- (I) will not be able decrypt the message, as it does not have the VAC.
- If (I) tried to tamper the message before delivering it to (V), (V) will be able to discover the tampering as this will change the f(N1).

The previous example can be generalized between any two VANET entities to be sure that the Man-in-the-middle-attack will not affect the dynamic key distribution protocol for VANETs.

### 3.6.3 Sybil Attack

Since every vehicle has a unique identity that cannot be tampered. Sybil attacks are counter-measured in the proposed protocol as per the following scenario:

- (I) will not be able to send a key request, as it requires knowing the ELP.
- If under any case (I) knows (V)'s ELP, it will send a key request to the RSU.

- The message reply containing the key will be encrypted with the VAC. In this case, (I) will not be able to make use of the generated key. Only (V) can use it, as it is the only holder of the VAC other than the CA.

### 3.6.4 Replay Attack

Nonces are used within the protocol messages. A nonce is a random or non-repeating parameter value that is included in data exchanged by a protocol, usually for the purpose of guaranteeing liveness and thus detecting and protecting against replay attacks. A nonce can be a timestamp, a visit counter on a Web page, or a special marker intended to limit or prevent the unauthorized replay or reproduction of a file. The following scenario describes how the proposed protocol responds to a replay attack:

- (V) will send a key request message concatenated with its nonce to the RSU.
- (I) will intercept (V)'s message.
- (I) will later resend the message to the RSU on behalf of (V).
- (V) can only decrypt the reply message containing the key, as no one knows the VAC other than (V) and the CA.
- When (V) decrypts the message, it will realize from the nonce that the message is an old message.
- (V) will ignore the message and request another key.

## 3.7 Maintained State Analysis

The proposed protocol relies on maintenance of state in the system. This section outlines what state needs to maintained by each entity in the system.

- **OBU**: each vehicle needs to maintain its ELP, and only the active key credentials instead of holding a large number of keys.
- **RSU**: needs to maintain a list of all vehicles passed by, and their next RSU (forwarding chain, see section 3.4) during the certificate lifespan.
- **RSU-Manager**: needs to maintain the information of RSUs under its authority, in addition to vehicles that acquired active certificates. Furthermore, it should keep a record of vehicles that moved between multiple RSU-managers to minimize delivery failure as possible.
- **CA**: should keep track of the blacklisted vehicles to be considered during the key issuance phase.

The proposed work decreased the required storage per vehicles to maintain a safer key storage process. On the other hand, the protocol increased the responsibility of RSUs and RSU managers who are considered secure infrastructure components to hold required information while their physical security is out of the scope of this paper.

TABLE 2
Comparison versus Related Work

| Related Work | Key Distribution | Revocation |
|---|---|---|
| Hubaux [5] | TPD stored keys | Global CRL Distribution |
| Papadimitratos [6] | TPD stored keys | Divide CRL to multiple lists and send |
| Aslam and Zou [13] | TPD stored keys | Geographical CRL Distribution |
| Zhang [14] | Dynamic Key Distribution | CRL Distribution |
| Wasef and Shen [15] | Dynamic Key Distribution | 2 CRLs (RSU, and OBU) |
| Hao et al. [17] | Group Key Distribution | CRL Distribution |
| Laberteux et al.[18] | TPD stored keys | V2V CRL Distribution |
| Nowatkowski [20] | TPD stored keys | Split CRL and send on different channels |
| Proposed Work | Dynamic Key Distribution | RSU scope CRL distribution |

### 3.8 Comparison versus Related Work

Previous approaches could not benefit from dynamic key distribution either because they depend on the TPD for keys storage [5-6] or because only the revocation issue is considered without trying to gain benefit of the available information. The proposed PKI infrastructure depends mainly on the idea of dynamic key distribution in [17, 18]. The vehicle requests a key which is issued by the CA. For the CA to send the certificate to the requesting vehicle, the location should be known at the sending time (RSU Manager and RSU). The certificate carries a parameter stating its lifetime, which will be the target for revocation. The revocation message should be sent before the certificate expires or it will be useless. Knowing the location of the vehicle at the time of requesting the key and the maximum speed limit for vehicles, it will be easy to target a limited region with the revocation message. This will help in optimizing the network overhead. Qualitative comparison versus related work is highlighted in Table 2. On the other hand, the proposed protocol expects the availability of secure reliable network infrastructure.

## 4. PERFORMANCE EVALUATION

Performance of the proposed protocol is measured analytically and through network simulation using ns-2 [21]. Simulation is performed using a simplified version of the protocol with single encryption for each message offering confidentiality as a proof of concept. Encryption was carried out using OpenSSL package [22]. However, the encryption overhead consumed less than 5% of the overall simulation time. For all simulation results, each experiment was repeated 10 times and the average is reported to be within a 95% confidence interval [23].

The foregoing analysis will consider vehicle mobility using two different mobility models as illustrated in Fig. 8. First, consider a vehicle moving inside a city with average speed limits and frequently changing directions. The Manhattan Grid Model or city model is used to represent this mobility scheme. Second, consider a vehicle moving on a highway, which most probably is in a fixed direction with high speed. The Random Way Point Model or Highway model is adopted for this mobility scheme. The mobility scheme trace files were generated using BonnMotion v1.4 generator [24, 25]. BonnMotion is Java software, which creates and analyses mobility schemes. It was originally developed by the Communication Systems group at the University of Bonn, Germany, where it served as a tool for the investigation of mobile ad hoc network characteristics. The mobility schemes are exported for use in ns-2.

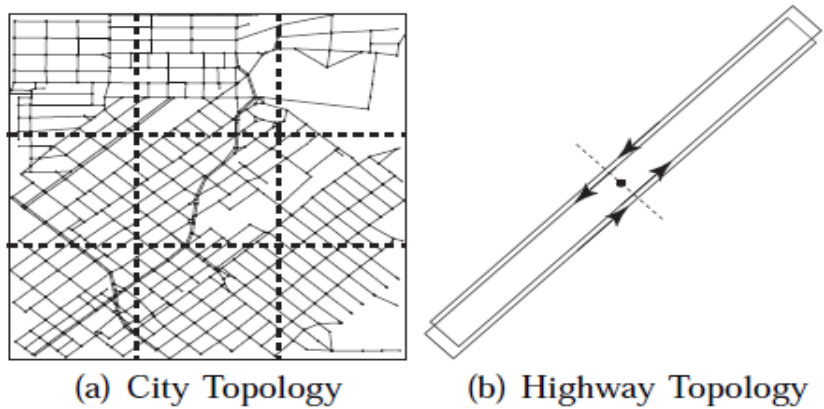

Figure 8. Mobility Models

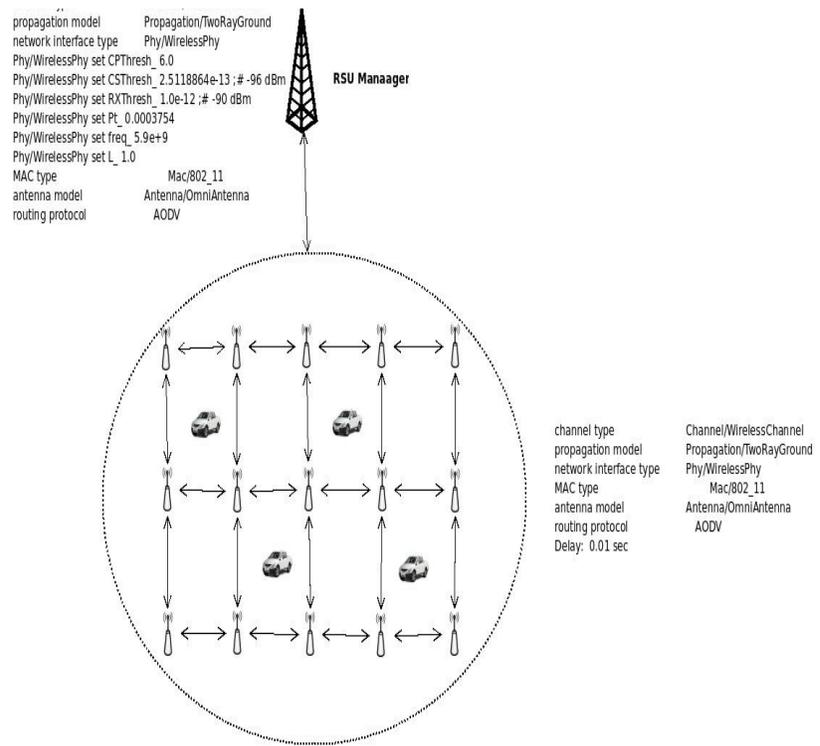

Figure 9. Network Simulation Setup

Fig. 9 illustrates setup for the simulation network. The RSU manager controls a group of RSUs that send messages to the passing vehicle according to the transmission range. When a vehicle moves from RSU-1 to RSU-2 and it receives messages at RSU-1. It will pass it to RSU-2 to be delivered to the vehicle as soft handover is assumed. The RSU manager's setup has some communication parameters to adjust the antenna used to send messages to longer distances.

The following three schemes were used during simulation for comparison. Broadcasting, Cooperative message authentication protocol (CMAP) [16], and Dynamic (proposed). The broadcasting scheme broadcasts the messages to all nodes (abbreviated hereafter as BRD). BRD uses $N$ messages for revocation [2]. There will be no effort for key distribution as the vehicle carries its own keys. In CMAP, the vehicle sends its position periodically [16]. The dynamic scheme is the protocol under consideration (abbreviated hereafter as DYN). Performance evaluation experiments include the following.

1. Effect of vehicle speed on number of revocation messages;
2. Effect of traffic density on number of revocation messages;
3. Effect of RSU manager control region on number of revocation messages;
4. Effect of node-to-node delay on effective message delivery;
5. Effect of traffic density on Packet delivery ratio; and
6. Effect of RSU manager count on number of revocation messages.

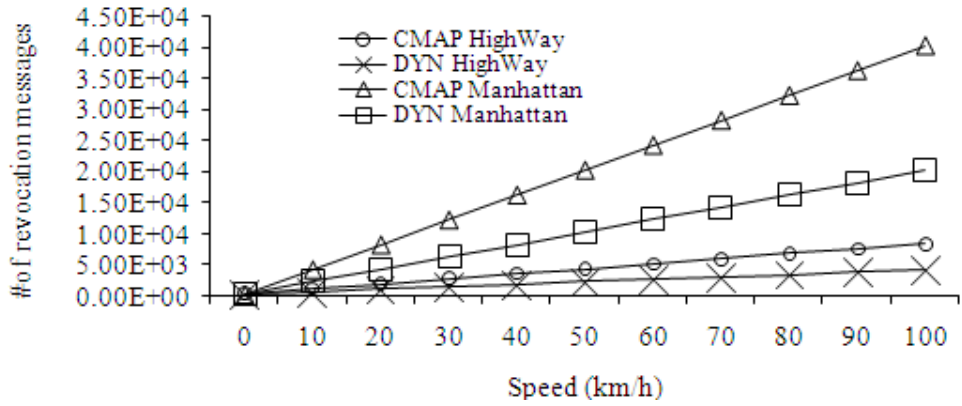

Figure 10. Number of messages to reach vehicles analytically [1]

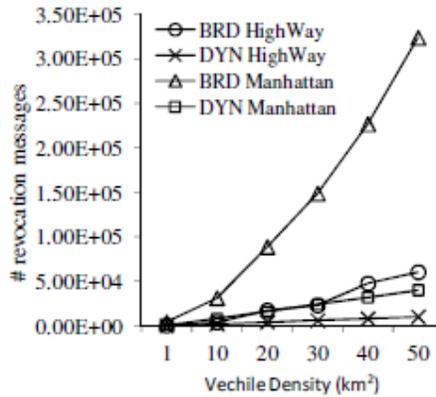

Figure 11. Effect of vehicle density on revocation

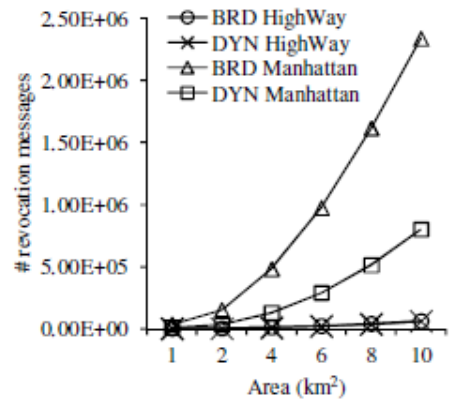

Figure 12. Effect of area on revocation

## 4.1. Effect of Vehicle Speed on Revocation Messages

Fig. 10 analytically shows the effect of changing the vehicle's speed and how this will affect the number of messages needed for revocation according to equations (1-3) in section 3.5. Manhattan Model assumes to have N=1000, v=80 km/h, d= 500 m, and l = 300 sec (5 min). Then $r$ = 6666.67 m, $m \approx 15$ messages, and $p$ = 1.5%. This means that we need to send 1.5 % of the messages sent in the broadcasting protocol as shown in Fig. 10.

Highway Model assumes to have N = 1000, v= 300km/h (an extreme), d = 1500m, and l = 900

sec. (15min). Then $r$ = 75000m, m = 51 messages, and $p$ = 5.1 %. This means that we need to send 5.1 % of the messages sent in the broadcasting protocol as shown in Fig. 10.

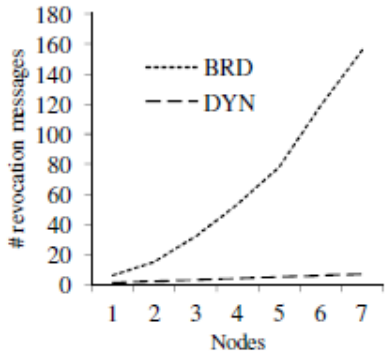
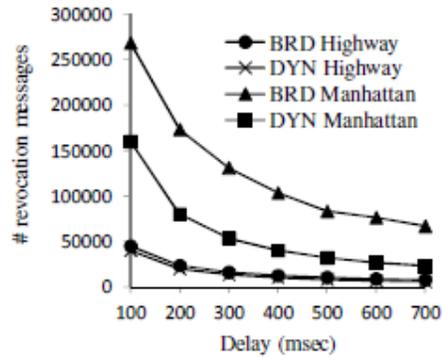

Figure 13. Messages for revocation analytically [1]       Figure 14. Effect of Node-to-Node delay

## 4.2. Scalability

For this experiment, the number of vehicles in a fixed area is increased to identify the number of messages needed for revocation. Fig. 11 illustrates the simulation results. For instance, with 30 vehicles the number of messages is 1.6% for Manhattan and 26% for Highway models of the number of broadcasting messages. To generalize, as the number of vehicles increases, the number of messages required for dynamic revocation is reduced as compared to previous approaches in both analytical and network simulation results.

## 4.3. Revocation Effort

The revocation effort in previous protocols was sizable as they depended on broadcasting in a specific region [6]. In previous protocols, the revocation messages were sent to the whole network either in a compressed form, or using multiple network channels, or to the close geographical region [15]. We will assume the close geographical region to be the circular area with radius $r$. We can make use of our knowledge of the vehicle area of presence to decrease our intended receivers to the minimum. As shown in Table 1, to reach the vehicle only $m$ messages are needed for DYN in comparison to $n$ messages for broadcasting. Fig.13 shows, analytically, how the number of messages sent to revoke a vehicle's key will differ between the dynamic and broadcasting protocols as the area increases. For simulation, the area was changed to investigate how this will affect the number of revocation messages. Fig. 12 shows the effect of increasing the area versus the number of messages required for revocation. Fig.12 shows no significant difference as the Highway model has a few OBUs and RSUs to make the difference between protocols. As the area increases, we can ensure that the number of messages required for dynamic revocation outperforms previous approaches through both analytical and network simulation results.

## 4.4. Delay

The node-to-node communication delay is a critical issue that has its effect on the whole network. As the delay increases the network performance dramatically decreases. In reference to Fig. 14, although the network performance decreases, DYN keeps its better performance compared to BRD. In both Manhattan and Highway models, as the delay between the nodes increases the number of messages passing through the network decreases in a way that gives the advantage to DYN against BRD. For example if the node-to-node delay is 300 msec, then Highway-DYN requires 84% of the revocation messages used by BRD. The DYN-Manhattan requires 40% of the revocation messages used by BRD for 300 msec delay.

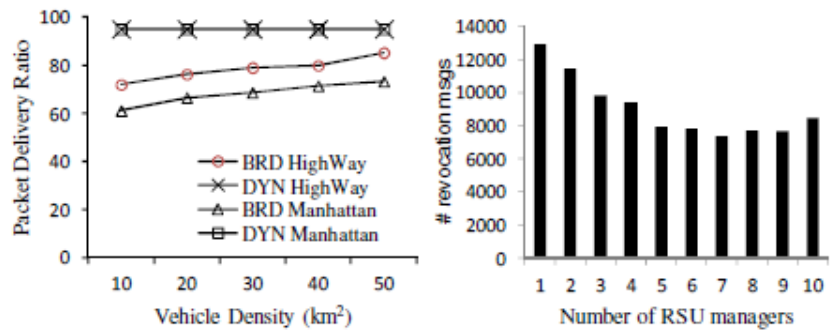

Figure 15. Packet delivery ratio        Figure 16. Effect of Number of RSU Managers

### 4.5. Packet Delivery Ratio

Fig. 15 depicts the efficiency of packet delivery when comparing DYN to BRD. It is clear that DYN achieves a constant delivery ratio regardless the traffic density. BRD starts with a low delivery ratio as it is supposed to reach out all vehicles in the system. On the other hand, DYN targets a small subset of the vehicles that are concerned with the revocation message achieving a higher delivery rate.

### 4.6. The effect of increasing the number of RSU Managers

The proposed protocol depends on the RSU manager in locating the RSUs and vehicles under its authority and using the best routes to deliver messages to the intended receivers with minimal number of sent messages. Fig 16 depicts how changing the number of RSU managers will affect the number of messages sent for revocation. The experiment was run for 100 sec. over an area of 4 km$^2$ (2 km x 2 km). The RSUs are divided equally among RSU managers depending on their regional location. When only 1 RSU manager is used, the revocation messages will be broadcasted to all RSUs which is equivalent to BRD's behavior. When using 2 RSU managers, the RSUs will be divided among both managers which will decrease the number of revocation messages although the vehicle did not cross the RSU manager administrative domain. If it crossed the domain, the messages will be passed to the RSUs controlled by the other manager as previously explained in section 3.4. This will decrease the number of sent revocation messages as the number of RSU manager's increases. At 8 RSU managers, the number of sent revocation messages would begin to increase again. This behavior is a result of the multiple divisions of the administrative domains. Starting from 8 RSU managers, the administrative domains will contain a fewer numbers of RSUs which will increase the probability for a vehicle to cross the administrative domain increasing the number of messages sent for revocation. This behavior will give an indication that the increase in the number of RSU managers along with a chain of RSUs will not enhance the performance all the way. The number of RSU managers, which will result in a small number of sent revocation messages, will vary according to the network size and structure.

## 5. CONCLUSION AND FUTURE WORK

In this paper, a lightweight dynamic PKI-based key distribution protocol for VANETs was proposed and thoroughly evaluated through analytical analysis and network simulation. The protocol reduces the role of TPD from being the main carrier and protector of the keys into carrying a key till the end of its lifetime. Vehicles will securely acquire the keys dynamically upon request. After the key is securely received, vehicles can communicate with other vehicles using the acquired key till its lifetime ends or it is revoked. If a vehicle's key was requested for revocation, the revocation message will be sent only to the vehicles that have a probability of communication with the revoked vehicle. Previous research could have sent to vehicles that would never have a probability to communicate with the revoked vehicle which in turns misuses the network resources. The proposed protocol efficiently reduces the revocation overhead and improves network utilization when compared to related work. The CA is assumed to be cloud-based but future work will consider cloud-based RSU managers as well. Furthermore, the current solution will be enhanced by eliminating the RSU manager chain and simulating the increase in the number of RSU managers to see how this will affect the number of revocation messages.

## Authors

**Ahmed H. Salem** received his Master of Science in Computer Engineering from the Arab Academy for Science, Technology, and Maritime Transport, Alexandria, Egypt, in 2012. He is currently a Ph.D . student at the Computer Science Department, Old Dominion University, VA, USA. His research interests include wireless networks, mobile computing, and network security.

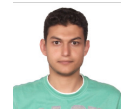

**Ayman Abdel-Hamid** is an associate professor in the College of Computing and Information Technology, Arab Academy for Science, Technology, and Maritime Transport (AASTMT), Alexandria, Egypt. In addition, he currently holds the position of Vice Dean for Postgraduate studies and scien tific research. He received his Ph.D. degree in Computer Science from Old Dominion University, VA, USA in May 2003. His research interests include mobile computing, network-layer mobility support, computer and network security, distributed systems, and computer networks. He is a member of IEEE, IEEE Computer Society, ACM, and ACM SIGCOMM.

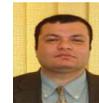

**Mohamad Abou El-Nasr** is a Professor of Computer Engineering, and Dean of Educational Affairs, Arab Academy for Science, Technology and Maritime Transport (AASTMT)—Alexandria, Egypt. He is also affiliated with Virginia Polytechnic Institute and State University where he works as an adjunct professor in the Bradley Department of Electrical and Computer Engineering - VTMENA program. He earne d both his Ph.D. and M.Sc. in Electrical and Computer Engineering in March 2003 and December 1999 respectively, from Georgia Institute of Technology, Atlanta GA, USA. His research interests include UWB systems, physical MAC layer issues in wireless networks, wireless sensor networks, cloud computing, e-health, and m-health. He is a Senior Member of IEEE Communications and Computer societies, and a member of ACM.

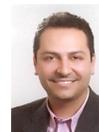